\documentstyle[12pt]{article}

\begin{document}

\author{Alexei V. Tkachenko \\
\it The James Franck Institute \\
\it The University of Chicago, Chicago , IL 60637}
\title{{\bf Effect of Chain Flexibility on Nematic--Smectic Transition }}
\maketitle

\begin{abstract}
The  theory of nematic--smectic phase transition in the system of
uniform semi--flexible chains with hard--core repulsion is presented. Both
the general density--functional formalism the tube--model
calculation show that the flexibility of the chains results in a strong 
{\em first--order} transition, in contrast to the common weak--cristallization
scenario of the nematic--smectic transition in rigid rods. The calculated spinodal volume
fraction of the uniform nematic phase and the period of the modulation
instability are consistent with recent experimental results.
\end{abstract}


\newpage 

\section{Introduction }

The nematic--smectic phase transition is among the most interesting
phenomena in  liquid crystals. This is one of a few examples of a {\em partial} breaking of  translational
symmetry: the system in smectic state has a layered structure, while the
translational symmetry along  the in--layer directions is preserved. It is
believed that normally the physical origin of the transition to the smectic
phase is the non--uniform architecture of the consistuent particles, such
that the nematic--smectic transition can be interpreted as a microphase
separation of different parts of the molecules. Nevertheless, it is
conceptually important to realize that a non--trivial structure of the
constituent objects is not a necessary condition for the formation of
smectic. Both analytical and numerical investigations show that an
entropy--driven nematic--to--smectic phase transition occurs even in the
system of uniform rigid rods \cite{rods}-\cite{pol}.

Recent experiments with rod--like viruses have confirmed the predictions of
such theories \cite{seth}, \cite{seth1}. They also showed that a
dramatic change in behavior takes place due to the finite flexibility of viruses. In
particular, the transition turns out to be of the strong first order rather
than of weak--crystallization type typical for most smectics. In addition, the volume fraction at the transition is
considerably higher in the case of flexible chains than for rods. The chains are strongly localized within the layers, so that the
periodicity of the smectic phase is nearly equal to the length of an individual
stretched chain, while in the stiff-rod case it is longer than the length of
a rod because of a  weaker localization. 

In a recent communication \cite{prl} we have proposed a theory based on the tube--model
for the description of the nematic--smectic phase transition in the
system of uniform semi--flexible chains with hard--core repulsion. This
model has been shown to capture all the experimentally--observed features of the transition.
In this paper we elaborate our approach by relating the  tube--model
calculation to a general density--functional formalism. It is shown
that the strong first order character of the transition and the
equivalence of the smectic period to the chain length follow from a
general form of the density functional of the system.

The strategy of the work is as follows. In Section 2 we review the 
density functional approach to the nematic--smectic transition. In Section 3
we derive the general form of the free energy of semi--flexible chains
expressed as a functional of the density of their mid--points. The breakdown
of the stability condition (positivity of the inverse structure factor)
yields the limit of stability of the nematic phase and the critical wave
vector of the modulation bifurcation. In Section 3 we obtain the parameters
of the density functional from the tube model of the nematic state. This
allows one to determine the parameters of the transition for various chain
lengths. The structure of the theory allows one to understand the difference
between the behavior of semi--flexible chains and that of rigid rods.

\section{Density--functional Approach to Nematic--Smectic Transition}

A powerful tool for the description of various types of crystallization, and
in particular, of the nematic--smectic phase transition is the
density--functional approach. Within such theories the free energy $F$ of
the system is parameterized by the one--particle density, $\rho \left( {\bf r}%
\right) $. The thermodynamic stability (or at least meta--stability) of the spatially uniform (e.g.,
nematic) phase with respect to density modulation is controlled by the sign
of the corresponding second variation of the free energy:

\begin{equation}
G^{-1}\left( {\bf q}\right) \equiv \frac 1{kT}\frac{\delta ^2F}{\delta \rho
_{{\bf q}}\delta \rho _{-{\bf q}}}>0  \label{con}
\end{equation}
Note that $G^{-1}\left( {\bf q}\right)$  has the meaning of the
inverse structure factor in the nematic phase ($G\left( {\bf q}\right)
\equiv \left\langle \delta \rho _{{\bf q}}\delta \rho _{-{\bf q}%
}\right\rangle $). The system becomes unstable with respect to the transition to a spatially modulated (smectic) state, when  inequality  (\ref{con}) is  violated at a  certain finite wave
vector, ${\bf q}_0$. In many cases the nematic--smectic transition can be
successfully described within the weak crystallization theory \cite{braz}. In this approach one assumes that near the transition, the local deviations of the density field are
dominated by one or by several critical density waves, i.e. 
\begin{equation}
\delta \rho \left( {\bf r}\right) \simeq \sum\limits_{\alpha =1}^n\rho
_\alpha \left( {\bf r}\right) \exp \left( {\rm i}{\bf q}_\alpha {\bf r}%
\right) {\bf +}c.c.
\end{equation}
Here ${\bf q}_\alpha $ are the  critical wave vectors. The spatial dependence of
the $n$--component order parameter $\rho _\alpha $ is supposed to be much
slower than the critical density modulation itself. 

The Landau--type
expansion of the free energy in terms of this order parameter is the essence
of the weak crystallization approach: 
\begin{equation}
F=\int \left( \frac \epsilon 2\delta \rho ^2+\frac \lambda 6\delta \rho
^3+\frac \gamma {24}\delta \rho ^4+\frac g2\sum\limits_{\alpha =1}^n\left|
\nabla \rho _\alpha \right| ^2\right) {\rm d}{\bf r}  \label{weak}
\end{equation}
The translational symmetry of the free energy demands that any term in the
above expansion is a combination of density waves with zero total wave
vector. Note that a large third--order term in this expansion would
result in a strong first order transition and thus would violate the
requirement of smallness of the order parameter. Therefore, the weak
crystallization theory is adequate only for the cases when the third--order
term is zero or small.
In particular, its applicability to the nematic--smectic phase transition is
normally justified by the broken rotational symmetry of the nematic phase.
Indeed, in the rotationally--symmetric case, the critical wave vectors would
lie on the sphere of radius $q_0$, and one could choose three critical
density waves with zero total wave vector. Therefore,  rotational
symmetry implies that there is a non--zero cubic term in expansion (\ref
{weak}). In the case of the nematic--smectic transition the degeneracy in
the orientation of the critical wave vector is lifted, and one cannot
construct a third--order combination of the critical density waves. Hence, the
weak crystallization theory  can usually be applied to this transition.

As an example, consider the transition to the smectic state in the system of
perfectly aligned hard rods. The density--functional theory, developed for
this system in reference \cite{rods}, shows that its basic physics can be
successfully described even in the second virial approximation:

\begin{equation}
\frac{F^{\left( rods\right) }}{kT}=\int {\rm d}{\bf r}\rho \left( {\bf r}%
\right) \ln \rho \left( {\bf r}\right) +\frac 12\int \int {\rm d}{\bf r}{\rm %
d}{\bf r}^{\prime }\rho \left( {\bf r}\right) v\left( {\bf r-r}^{\prime
}\right) \rho \left( {\bf r}^{\prime }\right)
\end{equation}
which yields the following simple expression for the inverse structure
factor: 
\begin{equation}
G^{-1}\left( q\right) \equiv \left( 1+8\Phi \frac{\sin qL}{qL}\right)
\label{str1}
\end{equation}
Here we are interested only in the wave vectors parallel to the nematic
director, ${\bf e}$.
The breakdown of the stability condition corresponds to the
transition to the spatially--modulated smectic phase. Since the instability
is dominated by a single density wave (with ${\bf q}\simeq \pm 1.5\pi {\bf e/%
}L$), the transition in the hard--rod system is of the {\em %
weak--crystallization} type. This means that the smectic modulation of hard
rods is weak near the transition point. As a result, the normal--to--layer
fluctuations of the rods are of the order of their length $L$, consistent
with the fact that the corresponding period $\lambda ^{*}\simeq 1.3L$
differs considerably from $L$. Although within the model of freely--rotating
rods \cite{pon}, \cite{pol} the phase transition turns out to be of the first order, it
is so weak that the transition can hardly  be distinguished experimentally
from a second--order one.

\subsection{Nematic--Smectic Transition of Semi--Flexible Chains}

Consider a system of worm--like uniform chains of length $L$ with hard core
diameter $D$. The single--chain Hamiltonian for a given conformation ${\bf r}%
(s)$ ($0<s<L$) has the form

\begin{equation}
H^{(0)}=kT\int\limits_0^L{\frac p4}\left( \frac{\partial {\bf t}_s}{\partial
s}\right) ^2{\rm d}s  \label{sing}
\end{equation}
Here $s$ is the coordinate along the chain contour and ${\bf t}_s=(\partial 
{\bf r}/\partial s)$ is a unit tangent vector. The parameter $p$ is the
persistence length of the chain in the isotropic phase. In the nematic
phase, however, there are two different scales which play the role of
persistence length \cite{we}. One of them is the typical distance between
''hairpins'', the points at which the tangent vector changes its direction by b$180^\circ$. This
length becomes exponentially large for high enough nematic order parameter
and we assume here that it exceeds the chain length, i. e. the conformations
of the chains are straight lines with only weak transverse fluctuations of
the tangent vector about the nematic director. The correlations of these
transverse fluctuations are determined by another length scale known as the  deflection
length, $\xi _{\bot }$. This scale, which is smaller then the bare
persistence length $p$, determines the thermodynamics of the system (the
free energy can be estimated as $kT$ per chain segment of length $\xi _{\bot
}$).

We now have to express the free energy as a functional of the density of the
chains' centers,

\begin{equation}
\rho \left( {\bf r}\right) =\sum\limits_{chains}\delta \left( {\bf r}(L/2)-%
{\bf r}\right)
\end{equation}
If the chains were infinite, the conformational free energy averaged over
scales larger then $\xi _{\bot }$ would be a local functional of the volume
fraction. For finite chains there is also a translational entropy
contribution and a correction due to finite density of chain ends: 
\begin{equation}
\frac{F^{\left( chains\right) }}{kT}=\int {\rm d}{\bf r}\left\{ \rho \left( 
{\bf r}\right) \ln \rho \left( {\bf r}\right) +f^{\left( con\right) }\left(
\phi \left( {\bf r}\right) \right) +\rho _{end}\left( {\bf r}\right) f^{(end)}
\left( \phi \left( {\bf r}\right) \right) \right\}  \label{free en}
\end{equation}
Because of the low density of  chain ends, $\rho _{end}$, their effect is
accounted for in the above expression by a term linear in $\rho _{end}$, coupled
to some local function of the volume fraction. Since the chains are
strongly stretched along the nematic director, $z$--axis, and  do
not form ''hairpins'', one can relate the local density of the ends $\rho
_{end}\left( z\right) $ and the local volume fraction of chains $\phi \left(
z\right) $ to the density field $\rho \left( z\right) $ (we will not
consider any fluctuations of these fields in the plane normal to $z$): 
\begin{equation}
\rho _{end}\left( z\right) =\rho \left( z{\bf +}L/2\right) +\rho \left( z%
{\bf -}L/2\right)  \label{end}
\end{equation}

\begin{equation}
\phi \left( z\right) =\frac{\pi D^2}4\int\limits_{-L/2}^{L/2}\rho \left( z%
{\bf +}s\right) {\rm d}s  \label{fi}
\end{equation}
Due to these non--trivial relationships between the three fields $\rho
\left( {\bf r}\right) $, $\phi \left( {\bf r}\right) $ and $\rho
_{end}\left( {\bf r}\right) $, the above local free energy functional is
becomes non--local when expressed in terms of a single density field, $\rho $. Such
non--local properties of the density functional are necessary for the
description of the  nematic--smectic transition. The semi--flexible chains are
unique in that sense that this functional has a well-defined general form
dictated by the property of locality on the mesoscopic scales (below $L$
and above $\xi _{\bot }$).

In order to study the stability of the nematic state consider the second
variation of the free energy at the fixed average volume fraction $\Phi $ (
and the corresponding density $\overline{\rho }=\Phi /V_0$, where $V_0\equiv
\pi D^2L/4$ is the volume of a single chain):

\begin{equation}
\frac{\delta F^{\left( chains\right) }}{kT}=\frac 12\int {\rm d}{\bf r}%
\left\{ V_0\frac{\delta \rho \left( {\bf r}\right) \delta \rho \left( {\bf r}%
\right) }\Phi +\left. \frac{\partial ^2f^{\left( con\right) }}{\partial \phi
^2}\right| _\Phi \delta \phi \left( {\bf r}\right) \delta \phi \left( {\bf r}%
\right) +2\left. \frac{\partial f^{(end)} }{\partial \phi }\right| _\Phi \delta
\rho _{end}\left( {\bf r}\right) \delta \phi \left( {\bf r}\right) \right\}
\end{equation}
Performing the Fourier transform of all the fields under consideration (
along the $z$--axis, neglecting their variations in other directions) and
expressing these fields in terms of the density deviations, $\delta \rho $,
we obtain the following diagonalized free energy: 
\begin{equation}
\frac{\delta F^{\left( chains\right) }}{kT}=\frac{V_0}\Phi \sum\limits_q%
\frac{\delta \rho _q\delta \rho _{-q}}2\left[ 1+\Lambda \left( \Phi \right)
\left( \frac L{\lambda \left( \Phi \right) }\frac{\left( 1-\cos qL\right) }{%
\left( qL\right) ^2}+\frac{\sin qL}{qL}\right) \right]
\end{equation}
Here the dimensionless parameter $\Lambda $ and the length $\lambda $ are
certain functions of the average volume fraction, but not of the chain
length or wave vector. Note the universality of the $\left( q,L\right) $%
--dependence of the inverse structure factor for semi--flexible chains: 
\begin{equation}
G^{-1}\left( q\right) \equiv \frac 1{kT}\frac{\delta ^2F^{\left(
chains\right) }}{\delta \rho _q\delta \rho _{-q}}=\frac{V_0}\Phi \left[
1+\Lambda \left( \Phi \right) \left( \frac L{\lambda \left( \Phi \right) }%
\frac{\left( 1-\cos qL\right) }{\left( qL\right) ^2}+\frac{\sin qL}{qL}%
\right) \right]  \label{St-fac}
\end{equation}
Before proceeding with the discussion of this general form of $G^{-1}$, we
present a simple model which yields the microscopic expressions for the
parameters $\Lambda \left( \Phi \right) $ and $\lambda \left( \Phi \right) $
appearing in (\ref{St-fac}).

\section{Tube--model Calculation}

The interactions of a chain with its neighbors can be modeled by confining
it in an effective tube. If the system is dense enough, it can be viewed as
a close--packed array of such tubes. This means that the average tube
diameter in the vicinity of some point ${\bf r}$, is $D/\sqrt{\phi \left( 
{\bf r}\right) }$, where $\phi \left( {\bf r}\right) $ is the local volume
fraction. Therefore, the allowed amplitude of fluctuations of a chain within
the tube is $\Delta =D\left( 1/\sqrt{\phi \left( {\bf r}\right) }-1\right) $%
. The corresponding conformational free energy of the chain can be evaluated
in the Gaussian approximation, with the confinement imposed by a fictitious
external field. The resulting value is $\frac 3{16}kT$ per correlation
length $\xi_\bot =\left( 2\Delta \right) ^{2/3}p^{1/3}$ \cite{od}. Note that one
can obtain this result, up to a numerical coefficient, from simple scaling
arguments, since $\xi $ can be identified with a typical contour length
between two reflections of the chain from the tube walls. We conclude that the
conformational free energy per unit volume is determined by the field $\phi
\left( {\bf r}\right) $: 
\begin{equation}
f^{\left( con\right) }\left( {\bf r}\right) =\frac 3{16}\frac{kT}{\xi_\bot \left(
\phi \right) }\frac{4\phi \left( {\bf r}\right) }{\pi D^2}
\end{equation}

We assume that the dominant  effect  responsible for the smectic ordering is that the 
{\em internal parts of the chains cannot occupy a ''shadow'' region in the
vicinity of a free end}. This is a particular realization of the {\em %
correlation hole effect}, \cite{DG}. In the extreme case of perfectly aligned rigid
rods the space behind the edge of one rod can be filled only by the
complementary end part of another one (see Figure 1a). However, in the
system of semi--flexible chains the size of the ''shadow'' region can be
reduced by appropriate readjustments of the conformations of the
neighboring chains, as shown in Figure 1b. The screening of the ''shadow''
region can be described by a {\em screening length} $l_s$. The conformational free energy penalty for the creation of the free space near the edge of
every chain is given by the product of the local transverse pressure $\Pi
_{\bot }=\phi \partial f^{\left( con\right) }/\partial \phi -f^{\left(
con\right) }$ and the typical volume of the ''shadow'' region $\gamma $$\pi $%
$l_sD^2/4$. Here $\gamma $ is a geometrical factor of order of unity.
Approximating the shape of a typical ''shadow'' region with a cone, one
obtains $\gamma =1/3$. The typical bending energy associated with the  distortion of chain contour needed for the screening of the ``shadow region''  is $kTD^2 p/l_s^3$ per chain involved. The screening length and the energy of the end defect is determined by the balance between the osmotic energy  penalty and the bending energy, i.e. they can be obtained  by minimization of the following free energy:
\begin{equation}f^{(end)}(l_s)=D^2\left(\frac{\gamma\pi l_s}{4}\Pi
_{\bot }+\frac{Zp}{l_s^3}kT\right)\end{equation}
Here $Z$ is the effective number of the distorted chains (fortunately, the final result is nearly insensitive to the choise  of this parameter). The minimal value of the above free energy is given by 
\begin{equation}f^{(end)}\simeq \frac{\pi\gamma D^2}{2}\Pi^{3/4}_{\bot }(kTp)^{1/4}\simeq\frac{\gamma kT}{8}\left( 1/\sqrt{\phi \left( z\right) }-1\right) ^{-5/4}
\end{equation}
The correspondong optimal screening length is 
\begin{equation}
l_s=2\left( 2D\right) ^{2/3}p^{1/3}\left( 1/\sqrt{\phi \left( z\right) }-1\right)^{5/12}\equiv 2l^{*}\left( 1/\sqrt{\phi \left( z\right) }-1\right)^{5/12}
\end{equation}
Here $l^*\equiv\left( 2D\right) ^{2/3}p^{1/3}$ is the fundamental length scale of the problem. Up to a $\phi$--dependent factor, it  determines both the screening length and the correlation length $\xi_\bot$ discussed earlier. For practical purposes,  the screening length can be taken roughly equal to  $l^*$. 
Note that our estimate of the energy of the end defect is somewhat different from the one described in ref. \cite{meyer}, which is based on elasticity theory. Elastic approach neglects the energy penalty associated with the  non-zero osmotic pressure $\Pi
_{\bot }$, which in our case turns out to be considerably stronger then next--order  elastic corrections.  
Nevertheless, the very notion of the  effective attractin between chain ends discussed in the present  paper is conceptually  close   to the one in ref. \cite{meyer}. 

Depending on the total chain length, one can distinguish between two
qualitatively different limiting regimes. If $L\ll l^{*}$, the screening
effect is not significant and the chains can be considered as rigid rods.
Here, we consider the opposite limit, $L\gg l^{*}$, when the total volume
fraction of the ''shadow'' regions is low and one can neglect their overlap
in the spatially--uniform nematic phase.

Summarizing the above calculation one can write down the total free energy
of the system: 
\[
F^{(chains)}=\frac{kT}{V_0}\int {\rm d}{\bf r}\left\{ V_0\rho \left(
z\right) \ln \rho \left( z\right) +\frac 3{16}\frac L{l^{*}}\frac{\phi
\left( z\right) }{\left( 1/\sqrt{\phi \left( z\right) }-1\right) ^{2/3}}%
\right. 
\]
\begin{equation}  \label{free1}
\left. +\frac \gamma {8}\frac{V_0\rho _{ends}\left( z\right) }{\left( 1/\sqrt{%
\phi \left( z\right) }-1\right) ^{5/4}} \right\}
\end{equation}
The first term accounts for the translational entropy of the chains, the
second one represents the bulk conformational free energy of infinitely long
chains and the last term is due to the end anomalies. One can easily verify
that our model free energy has the general form obtained in the
previous section, (\ref{free en}). There is a one--to--one correspondence
between the three discussed contributions to the free energy and the three
terms in the following expression for the inverse structure factor in
 the uniform nematic phase, as a function of the average volume fraction $\Phi $: 
\begin{equation}  \label{str}
G^{-1}\left( q\right) \equiv \frac 1{kT}\frac{\delta ^2F}{\delta \rho
_q\delta \rho _{-q}}=\frac{V_0}\Phi \left( 1+\frac 5{16}\frac 1{\Phi \left(
1/\sqrt{\Phi }-1\right) ^{9/4}}\left[ \frac{2L}{3l_s(\Phi)} \frac{%
\left( 1-\cos qL\right) }{\left( qL\right) ^2}+\gamma\sqrt{\Phi} \frac{\sin qL}{qL}%
\right] \right)
\end{equation}

This expression has exactly the same general structure as has been derived
in the previous section. The uniform nematic is stable (or at least
metastable) with respect to the transition to the spatially--modulated
(smectic) state as long as the calculated inverse structure factor is
positive. The end effect contribution is the only term in expression (\ref
{str}) which may be negative (due to the sign--changing factor $\sin \left(
qL\right) /qL$). For most wave vectors, however, this term can not change
the overall sign of the structure factor because of the dominant positive
conformational contribution, which contains the large factor $L/l_s$. This
is related to the fact that the ''shadow'' regions are {\em screened} in the
many--chain system, and the corresponding end effect is just a small
correction to the conformational free energy. This correction is important
only in the vicinity of the zeros of the expression $\left( 1-\cos qL\right)
/\left( qL\right) ^2$, which determines the $q$--dependence of the bulk
conformational contribution to the inverse structure factor.

We conclude that the modulation instability in the system is expected only
for nearly $L$--periodic density waves, which are the zero modes of the
conformational term. One can expand the inverse structure factor in the
vicinity of such wave vectors, $2\pi n/L$ ($ n=\pm 1,\pm 2,...$): 
\[
\left. G^{-1}\left( \delta q\right) \right| _{\delta q=q-2\pi n/L}=\frac{V_0}%
\Phi \left( 1+\frac 5{16}\frac 1{\Phi \left( 1/\sqrt{\Phi }-1\right)
^{9/4}}\left[\frac{L}{3l_s}\left( \frac{\delta qL}{2\pi n}%
\right) ^2+\gamma{\sqrt {\Phi}} \left( \frac{\delta qL}{2\pi n}\right) \right] \right) 
\]
Its minima are the candidates for the critical wave vector of the modulation
instability: 
\begin{equation}  \label{fam}
\begin{array}{cc}
q_n^{*}=\frac{2\pi n}L\left( 1-3\sqrt{\Phi} \gamma \frac{l_s(\Phi)}L\right) , & n=\pm
1,\pm 2,...
\end{array}
\end{equation}
The control parameter $\Phi $ at which $G^{-1}\left( q_n^{*}\right) =0$
turns out to be independent of $n$ (up to a cut--off $n_{\max }\sim L/l^{*}$
where the small--$\delta q$ expansion becomes inadequate): 
\begin{equation}  \label{volfr}
\Phi ^{*}=\left( 1+\left( \frac{15\gamma ^2l^{*}}{16 L}\right)
^{6/11}\right) ^{-2}\simeq\left( 1+\gamma\sqrt{\frac{l^{*}}{ L}}\right) ^{-2} 
\end{equation}

The existence of the family of critical wave vectors (\ref{fam}) which
differ only by integer multiplier is the signature of{\em \ the first--order
phase transition }to the smectic state. Indeed, unlike the case of a {\em %
single} dominating density wave typical for most smectics (e.g. for rigid
rods), this {\em degeneracy} enables one to compose the third--order
combinations of the critical modes ($\psi \left( q_n^{*}\right) \psi \left(
q_m^{*}\right) \psi \left( q_l^{*}\right) $, $n+m+l=0$), which contribute to
the term $\left( \delta \psi \left( z\right) \right) ^3$ in the
density--deviation expansion of the free energy, (\ref{weak}). The non--zero
cubic term in the Landau expansion is known to result in a first--order
transition.

The common period of all the critical modes is 
\begin{equation}
\lambda ^{*}=L+3\sqrt{\Phi^*} \gamma {l^s(\Phi^*)}  \label{lamb}
\end{equation}
In a general case this critical period can differ from that of the
equilibrium modulated phase. Nevertheless, the above result suggests that
the period of the smectic $\lambda $ nearly coincides with the chain--length 
$L$, and that the small correction $\lambda -L$ is of order of $l_s\simeq l^{*}$. This
correction determines both the typical gap between well--formed smectic
layers and the typical longitudinal fluctuations of the chains in these
layers. This is consistent with the observations reported in ref. \cite{seth}%
, for which the calculated length $l^{*}\simeq 100nm$
is of the order of the measured correction to the period $\lambda -L\simeq
50nm$. The effective hard core diameter in the experiments can be estimated as the inter--chain separation at which the electrostatic repulsion becomes of the  order of $kT$. This length  depends on the ionic strength, so it was possible to change it independently of the particle density. Since $l^*\equiv(2D)^{2/3}p^{1/3}$ depends on the hard core diameter $D$, the measured points on the phase diagram (concentration--ionic strength) can be transformed to  the coordinates of our theory($\Phi$ -- $L/l^*$). The calculated spinodal volume fraction, Eq.(\ref{volfr}) is in an 
agreement with the  experimental value, which is about $0.75$ for $L/l^{*}$ in
the range from $4$ to $10$. Note that the chain--length dependence of the
critical volume fraction is rather weak, as shown in Figure 2.

\section{Discussion and Conclusions}

We now compare the results obtained for semi--flexible chains with those for
perfectly--aligned rigid rods. Consider the inverse structure factor of
perfectly--aligned rigid rods, Eq. (\ref{str1}): 

$$
G^{-1}\left( q\right) \sim \left( 1+8\Phi \frac{\sin qL}{qL}\right)
$$
The first term here is due to translational entropy and the second one is
the excluded volume contribution, which is essentially the ''shadow'' region
end effect. Naturally, the rigid--rod  structure factor  does not
contain the conformational contribution, which dominates the similar
expression for semi--flexible chains. Hence, the ''shadow'' region effect,
which drives for the nematic--to--smectic transition is no longer a small
correction. As a result, the transition in the system of rods takes
place at a lower volume fraction, which is $\Phi ^{*}\simeq 0.57$ within the
second virial approximation, Eq. (\ref{str1}). Taking the higher virial
terms into account changes this value to $0.36$. If the rods are freely rotating, the critical volume fraction depends on the length--to--diameter ratio. For long enough chains ($L/D\gg 10$) the transition volume fraction reachs a ``universal'' value $\Phi\simeq 0.46$ \cite{pon}, \cite{pol}. In the case of considerable flexibility of the
''molecules'', the critical (spinodal) volume fraction given by expression (%
\ref{volfr}) is not constant even for high length--to--diameter ratio, since the relavant parameter here is $L/l^*$, rather than $L/D$ (see
figure 2).

The typical behavior of the inverse structure factors for rigid rods, Eq. (%
\ref{str1}), and semi--flexible chains, Eq. (\ref{str}), are depicted in
figure 3. Unlike the  case of semi--flexible chains ($L\gg l^{*}$), in which the $%
q^{*}$--degeneracy of the bifurcation point results in a strong first--order
transition, the deepest minimum of the rigid--rod inverse structure factor
determines a {\em single} critical wave vector of the modulation
instability. This explains why the nematic--smectic transition for rigid
rods is much softer then for semi--flexible chains.

Another important implication of the theory is the effect of polydispersity.
Since the inverse structure factor describes the effective two--body
interactions, expressions (\ref{str}) and (\ref{str1}) can be extended to
the polydisperse case by replacing a single parameter $L$ with the mean
length of two interacting particles $\left( L_1+L_2\right) /2$, and
averaging over the distribution of lengths. The polydispersity acts against
the modulation instability, because it reduces the depth of minima of the
inverse structure factor. Hence, the critical volume fraction is expected to
increase with polydispersity for both  chains and rods. However,  while the smectic phase can form in systems of  rods with quite
broad distribution of lengths \cite{slu}, the typical deviation of the chain--length,
which completely suppresses the transition in systems of chains  is of the order of $l^{*}$. Thus, 
in order to observe the smectic phase in the system of chains the
distribution of lengths has to be very narrow.

It should be noted that although we have shown that the above behavior
follows from a very general form of the density functional for 
semi--flexible chains, our formalism cannot be directly applied to the case
of long--range (hexagonal) in--layer structure. In that case, a chain end
becomes a real topological defect, and there is no reason to expect that the
corresponding energy penalty is finite, i. e., that the corresponding
contribution to the  total free energy, Eq. (\ref{free en}), is linear in
the chain--end density.

In summary, the theory of nematic--smectic phase transition for uniform
semi--flexible chains with hard--core repulsion has been developed. Similarly to
the case of rigid rods, the transition is driven by the ''shadow'' region
end effect. The difference is that due to the finite flexibility of the
chains this effect is screened and the size of the empty space near a free
end is limited by the screening length $l_s\simeq l^{*}\ll L$. The presence of
''shadow'' regions yields just a small correction to the conformational free
energy which  stabilizes the spatially uniform nematic state. As a
result, the spinodal volume fraction for flexible chains is much higher than
for rigid rods. This trend is consistent with the experiments as well as with the recent  theory  of weakly flexible rods \cite{vds}. The modulation instability of the nematic state can appear
only for nearly $L$--periodic density waves, which are the soft modes of the
bulk conformational free energy. Therefore, unlike the rigid--rod case, the
period of the smectic phase near the transition point almost coincides with
the chain length $L$. Another important difference is that in the case of
semi--flexible chains the point of modulation instability is highly
degenerate in the critical wave vector resulting in a strong first--order phase
transition. The theory also implies that the flexibility of the chains
results in higher sensitivity of the transition to polydispersity. The
agreement of the theory with existing experimental data confirms that it
captures the basic physics of the phenomenon.

\medskip\ 

{\bf Acknowledgement}

\medskip\ 

The author thanks Y. Rabin, T. Witten, E. Gurovich for valuable discussions of this work. This research was supported in part  by MRSEC program of the National Science Foundation through grant NSF DMR 9528957 and by  grants from the Israeli Academy of Science
and Humanities and the Research Authority of the Bar--Ilan University.

\newpage\

\newpage\ 

Figure Captions.

\begin{description}
\item  Figure 1. The ''shadow'' region (dashed) in the case of rigid rods
(a) and semi--flexible chains (b). Note the {\em screening} of the
''shadow'' by neighboring chains in the latter case.

\item  Figure 2. Spinodal volume fraction of uniform nematic $\Phi ^{*}$ as
a function of reduced chain length $L/l^{*}$. Solid line corresponds to the
theoretical result obtained for semi--flexible chains ($L/l^{*}\gg 1$), and
the dashed one is an interpolation of the crossover to the rigid--rod limit (%
$L/l^{*}\ll 1$). The geometrical factor $\gamma $ is taken to be $1/3$, as
is suggested in the text. Experimental points (diamonds) are taken from ref.
[5].

\item  Figure 3. Typical inverse structure factors (in arb. units) of the nematic phase for
rigid rods (dashed line ) and semi--flexible chains (solid line).
\end{description}

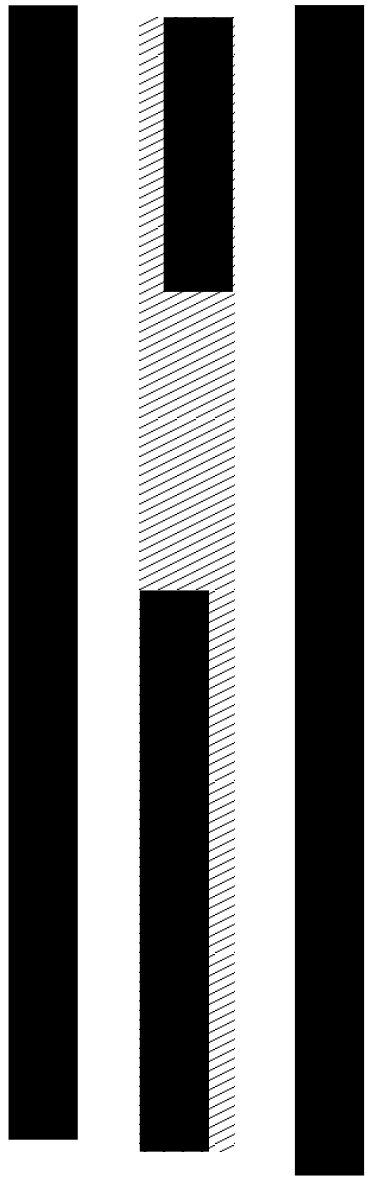
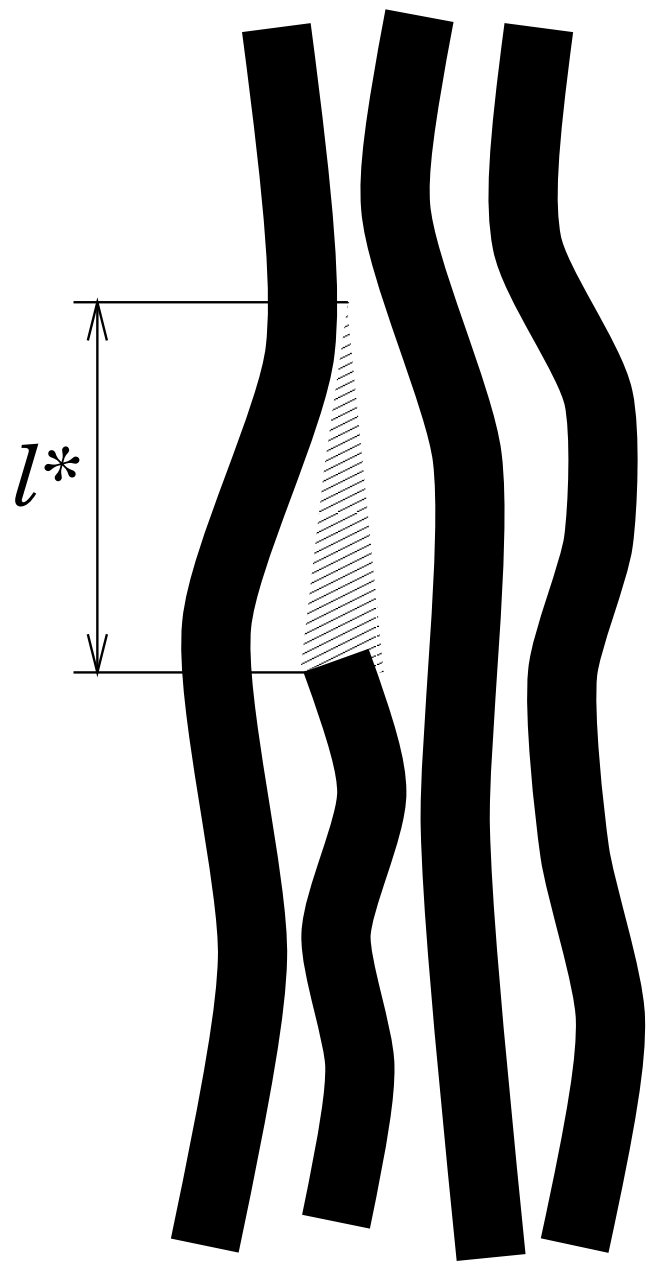
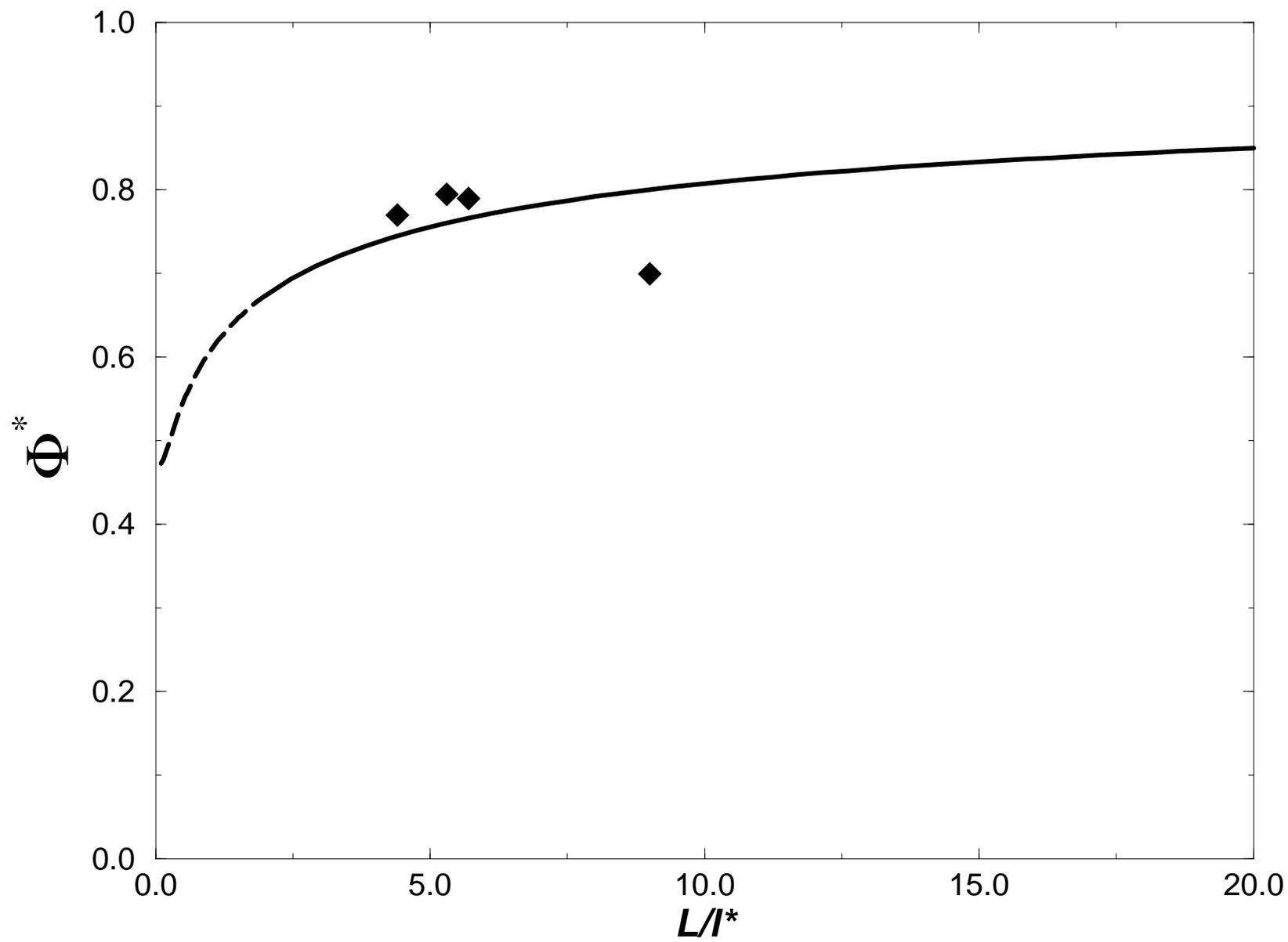
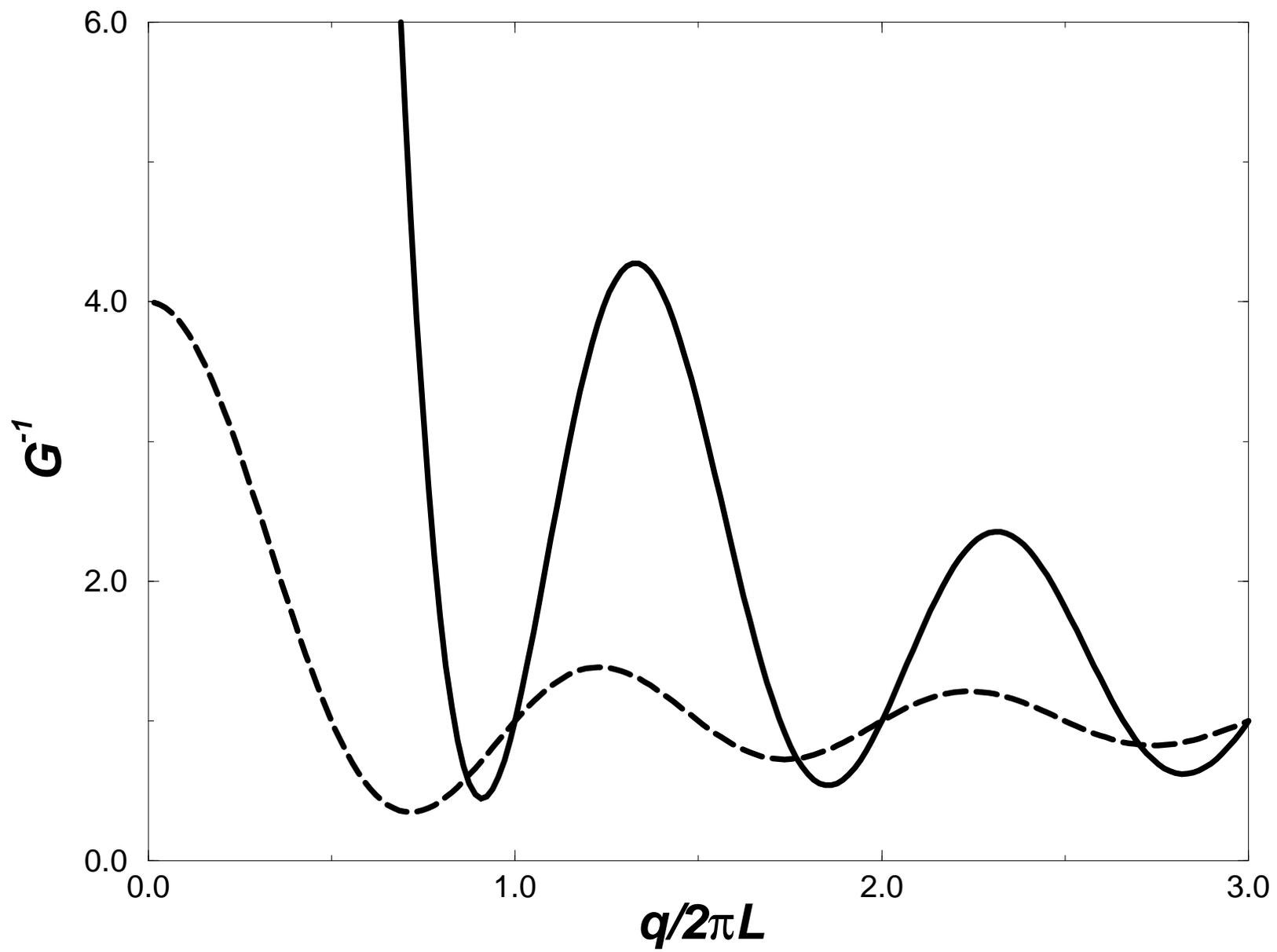
\end{document}